\documentstyle[12pt,epsfig,aps]{revtex}
\newlength{\mpclength}

\begin{document}
\title{What determines the $K^-$ multiplicity at energies around~1-2~AGeV?
}
\author{Ch. Hartnack$^1$ H. Oeschler$^2$ and
J. Aichelin$^1$ }
\address{
$^1$SUBATECH,
Laboratoire de Physique Subatomique et des
Technologies Associ\'ees \\University of Nantes - IN2P3/CNRS - Ecole des Mines
de Nantes \\
4 rue Alfred Kastler, F-44072 Nantes, Cedex 03, France\\
$^2$Institut f\"ur Kernphysik, Darmstadt University of Technology,
D-64298 Darmstadt, Germany\\}

\author{\begin{quote}
\begin{abstract}
 {In heavy ion reactions at energies around 1-2 AGeV 
the measured $K^-$ yields appear rather high as compared to pp collisions
as shown by the KaoS collaboration.
Employing Quantum Molecular Dynamics (IQMD)\cite{iqmd} simulations, we show that this is caused by the fact 
that the dominant production channel is not $BB\rightarrow BBK^+K^-$ 
but the mesonic $\Lambda(\Sigma) \pi \rightarrow K^-B$ reaction. Because 
the $\Lambda $ ($\Sigma$) stem from the reaction 
$BB \rightarrow \Lambda(\Sigma) K^+ B$, the $K^+$ and the $K^-$ yield 
are strongly correlated, i.e.~the $K^-/K^+$ ratio occurs to be nearly
independent of the impact parameter as found experimentally.   $K^-$ are
continuously produced but also very quickly reabsorbed leading to an almost
identical rate for production and reabsorption.
The final  $K^-$ yield is strongly influenced by the $K^+N$ (due to their 
production via the $\Lambda(\Sigma)$) but very little by the $K^-$N potential.}    
\end{abstract}
\end{quote}}
\date{\today}
\maketitle

 A while  ago the KaoS collaboration has published results on the 
 $K^-$ and $K^+$ production in Ni+Ni reactions at 1.8 AGeV  and 1.0 AGeV, 
 respectively, \cite{Barth} which came as a surprise:  As a 
 function of the available energy, i.e.~$\sqrt{s} - \sqrt{s_{\rm threshold}}$ 
 (where $\sqrt{s_{\rm threshold}}$ is 2.548 GeV for the $K^+$ 
via $pp \rightarrow \Lambda K^+p$ and 2.870 GeV 
 for the $K^-$ via $pp\rightarrow ppK^+  K^-$)
the number of  $K^-$  produced equals that of $K^+$ although in pp reactions 
close to threshold the cross section for $K^+$ production 
is orders of magnitude higher than that for $K^-$ production. 
 In addition, the $K^-$ have a high probability for absorption
via $K^- N \rightarrow \Lambda  \pi$, whereas the $K^+$ cannot get 
reabsorbed due to its $\bar s$ content.
Even more astonishing was the experimental finding that 
at incident energies of 1.8 and 1.93 AGeV
the $K^-$ and $K^+$ multiplicities exhibit the same impact 
parameter dependence \cite{Barth,Menzel,sqm} although 
the $K^+$ production is above the respective $NN$ threshold while the
$K^-$ production is far below. Equal centrality dependence for  $K^+$ and $K^-$
was also found at AGS energies \cite{Ahle}.
All these observations have triggered a lot of activities 
\cite{schaffi} - \cite{ho_s2000}.

It is the aim of this Letter to show that these observations have 
a simple explanation. For 
this purpose we use  IQMD simulations. The details of the IQMD approach have 
been published elsewhere \cite{iqmd}.
Here we have introduced in our standard simulation program
a (density-dependent) $KN$ potential.  We use the results of the
relativistic mean field (RMF) calculation of Schaffner \cite{schaf} 
which gives, as shown in this paper, the same result as more sophisticated
approaches like the chiral perturbation theory or the Nambu-Jona-Lasinio
Lagragian. The relativisitc mean field shifts the masses of the particles 
in the 
medium. Because this mass shift is applied to the phase space as well as to 
the flow, detailed balance in the production cross sections is conserved.
We have supplemented as well our calculation by all revelant cross sections
for kaon production and annihilation. The added cross sections are 
either parametrisations of the experimentally measured elementary cross 
sections or are based on one-boson-exchange calculations 
if measured cross sections are not available. The kaons are treated
perturbatively. This means that in each collision with a sufficient center of
mass energy there is a kaon produced with a probability given by the relative
ratio $\sigma(NN\rightarrow K^+ +X)/\sigma_{tot}$. The final momenta of the 
nucleons, however, are calculated under the assumption that no kaon has been
produced. If the kaon-nucleon  potentials are switched on we assume that the 
in-medium cross section and the free cross section agree for the same 
relative momentum between the scattering partners.    
Recently Lutz and al. \cite{lutz} as well as Schaffner et al.
\cite{schaffi,scha} have calculated some of the relevant in medium cross 
sections using a coupled channel approach for the t-Matrix. However, 
their results differ considerable and the former one is not adapted yet 
to simulation programs using spectral functions instead of quasi particles.

Therefore and because the physics discussed in this Letter depends only little 
on the transition matrix elements but predominantely on the mass difference 
between entrance and exit channel which does not change decisively in the 
medium we do not employ these improved cross sections here. It is suffcient 
to vary the experimental cross section by an artifical multiplication factor 
to see whether how the physics depends on the cross section. 

The strange baryons $\Sigma$ and $\Lambda$ are treated as one particle and in the text we use 
only $\Lambda$ which stands for both particles. The potential $U_{\Lambda N}$ 
is taken as 2/3 of $U_{NN}$. This yields a very good description of the $K^+$
and $\Lambda$ production in these heavy ion collisions \cite{ikaon}.
Here we concentrate on an understanding of the $K^-$ production. 
We have performed the calculations for 
the reactions Au+Au  and C+C at 1.5 AGeV incident energy. This energy has 
been chosen because data on $K^-$ and $K^+$ for three different 
systems are available \cite{sqm} or will soon become available. 

\begin{figure}[b]
\psfig{figure=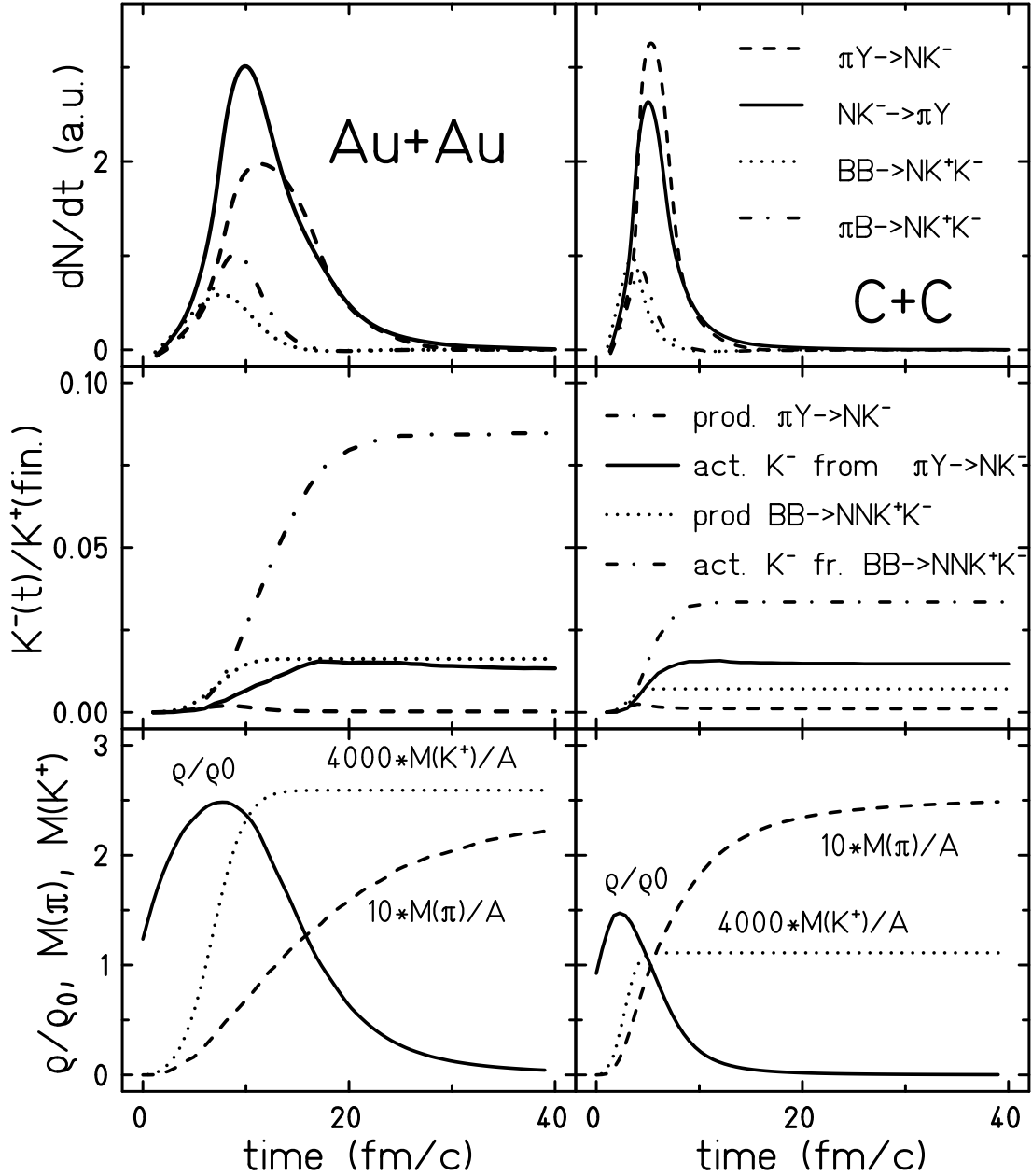,width=0.9\textwidth}
\caption{
Time evolution of rates and multiplicities for central collisions
of Au+Au (left) and C+C reactions at 1.5 AGeV incident energy.
Top panel: rates for the production and absorption channels for $K^-$.
Middle panel: Integrated number of produced $K^-$  and
the actual number of $K^-$ (production minus absorption) present at time $t$ for
different $K^-$ production channels (normalized to the
final number of $K^+$).
Bottom panel: Number of free pions/$A$ and of $K^+$/$A$
(with $A$ the mass number of one collision partners) and the central density
as a function of time.}
\label{channel}
\end{figure}

We start out by investigating in which channels the $K^-$ are 
produced and absorbed in the course of the reaction. Figure \ref{channel} 
displays the production and absorption rates (top panel) of the $K^-$ as well
as the integrated number of produced $K^-$  and
the actual number of $K^-$ present in the system as a function of time (middle
panel) for two production channels  
$\pi \Lambda \rightarrow K^-  B$ and  
$BB \rightarrow BBK^+  K^-$
where $B$ is either a nucleon or a $\Delta $. The third production channel
$\pi B \rightarrow B K^+  K^-$ has qualitatively the same structure as the BB
channel. On the left (right) hand side we display the results for central 
reactions 
($b = 0$) at 1.5 AGeV for Au+Au (C+C) collisions. 
For reasons which we will discuss later, the $K^-$ yields are divided by the 
final number of $K^+$. 
In the bottom panel the density, the number of free pions 
and the number of $K^+$ (both divided by the mass number $A$ of one of the
nuclei) are given as a  function 
of time. It has already been pointed out in ref. \cite{cassing,koli} 
that $K^-$ can also be produced in the 
strangeness-exchange channel
$\pi\Lambda \rightarrow K^-B $ channel.
Our calculation Fig.~\ref{channel} shows that for
both targets $\pi\Lambda \rightarrow K^-B $ is the dominant $K^-$ 
production channel.
The $K^-$ from the $BB$ channel are produced in the high-density 
phase and earlier than those from the $\pi \Lambda$ channel because 
the $\Lambda$ have to be produced first in $BB \rightarrow \Lambda K^+  B$ 
collisions. At a later stage of the collisions the available energy in the 
$BB$ and $\pi B$ channels is not sufficient anymore to create kaon pairs. 
Produced early, the 
$K^-$ from the BB and $\pi B$ channels have a higher chance to be reabsorbed and 
finally almost all (more than 90\%) of them have disappeared.
Therefore, the observed $K^-$ are essentially from the strangeness-exchange
process. 
In this channel the rates of production and absorption after 15 fm/c are 
about equal in the Au+Au system and very similar in the C+C system.
In general, absorption of $K^-$ is very high:
in C+C collisions 
about 50\%  survive whereas in the Au+Au collisions this 
is only the case for less than 20\% as can be seen from the middle part of
Fig.~\ref{channel}. We like to note that the final number of $K^-$ divided
by the $K^+$ is equal in both systems despite of the fact the many more
$K^-$ are produced.

\begin{figure}
\psfig{figure=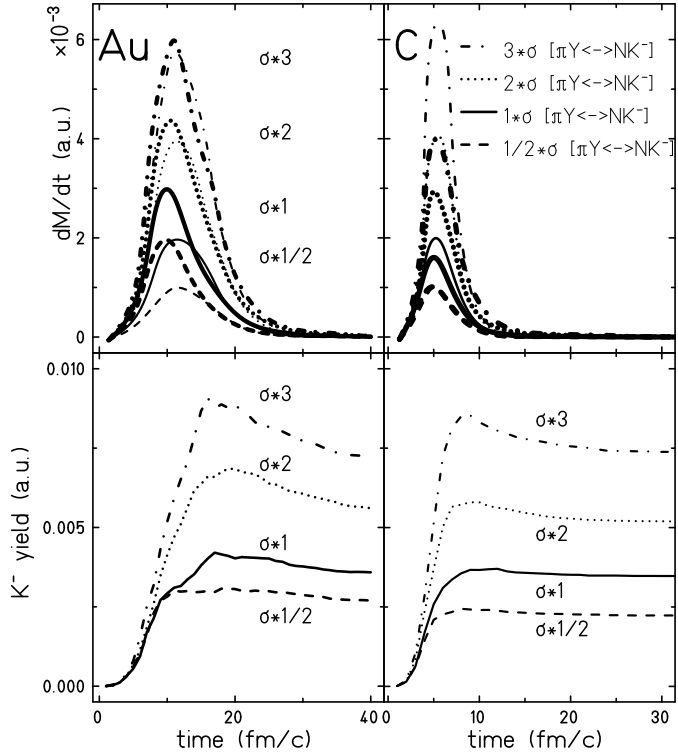,width=0.55\textwidth}
\caption{Top panel: The rates for creation (thin lines) and absorption 
(thick lines) of $K^-$.
Bottom panel:
The net number of $K^-$ in the system. The effective cross sections we used in the
calculation are denoted by $\sigma_{eff} = n \sigma_{exp}$ with
n being a multiplication factor. On the left (right) hand side central
Au+Au (C+C) reactions are shown, both at 1.5 AGeV.}
\label{sigma}
\end{figure}

Next we study the balance between creation and absorption of $K^-$.
In the upper part of Fig.~\ref{sigma} we plot 
the rate of absorption and production of the $K^-$ in the 
dominant $\Lambda \pi $ channel as a function of time. 
A large difference in the rates is seen at the moment when the system 
is very dense. There the $K^-$ are produced via $K^-K^+$ pair production
as already shown in Fig.~\ref{channel}. 
Later, during
expansion,  the  rates for absorption and production become very similar.
Finally the rates separate again  but now the rate
for $K^-$ absorption dominates. The $K^-$ production
approaches zero because for this endothermic reaction the energy in a
$\Lambda\pi$ collision is not sufficient anymore but the exothermic
inverse reaction can still continue. 
This difference of the rates in favor of the 
exothermic reaction 
is a very general phenomenon in expanding systems where the locally 
available energy decreases with time \cite{pal}. 

The fact that the rates in both directions are almost identical testifies that
the system has reached an equilibrium. This may be a thermal equilibrium 
which is characterized by the fact that the particle yields depend exclusively 
on the temperature and the chemical potential. It may be as well a steady state
which occurs for example if creation and absorption are strongly
connected by a very short life time. 

In view of this ambiguity it is useful to test whether the
system has reached thermal equilibrium: If it is in thermal equilibrium
an (artificial) increase of the cross sections does not change the particle 
number ratios (and hence also not the number $K^-$). It only 
brings the system faster to equilibrium. Therefore we can test whether thermal
equilibrium 
is obtained by multiplying the cross section in the $\pi \Lambda $ 
channel by a constant factor $n$.  From the top part of 
Fig.~\ref{sigma} we see that both, production and absorption,
increase with $n$, but differently. 
The net numbers are given in the lower part of 
Fig.~\ref{sigma} showing that a larger cross section produces more $K^-$. 
Thus we can conclude that for $n$=1 the system is not yet in (local) 
thermal equilibrium. However, changing the cross section by a factor of two 
changes the number of $K^-$ by 56 (42)\% only in the Au+Au (C+C) system 
(and not by a factor of two). This interpretation is supported by the
observation that only few of the $\Lambda$ make a $K^-$ and that 
only a negligible number of those $\Lambda$, which are produced by $K^-N$ 
collisions, produce another time a $K^-$. Therefore there are too few 
collisions in this channel to produce an equilibrium. 

Thus in the present situation we observe a steady state. Its origin 
is easy to understand if one  realizes that close to the threshold 
due to flow and phase space the cross section of the dominant channel,
$\pi \Lambda \rightarrow K^- N$, 
is very small as compared to that of the inverse reaction. Rarely a $\Lambda$
produces a $K^-$. If it does, the mean free path
and hence the mean lifetime of the $K^-$ is short. 
It will be destroyed  shortly after its creation with the consequence that 
the rate of production and annihilation are identical. 
Only close to the surface the $K^-$ has a chance to
escape. Thus we find here a system in equilibrium (in the sense that both rates
are identical) but not in thermal equilibrium (in the sense that particle ratios
are given by temperature, chemical potential). 

\begin{figure}
\psfig{figure=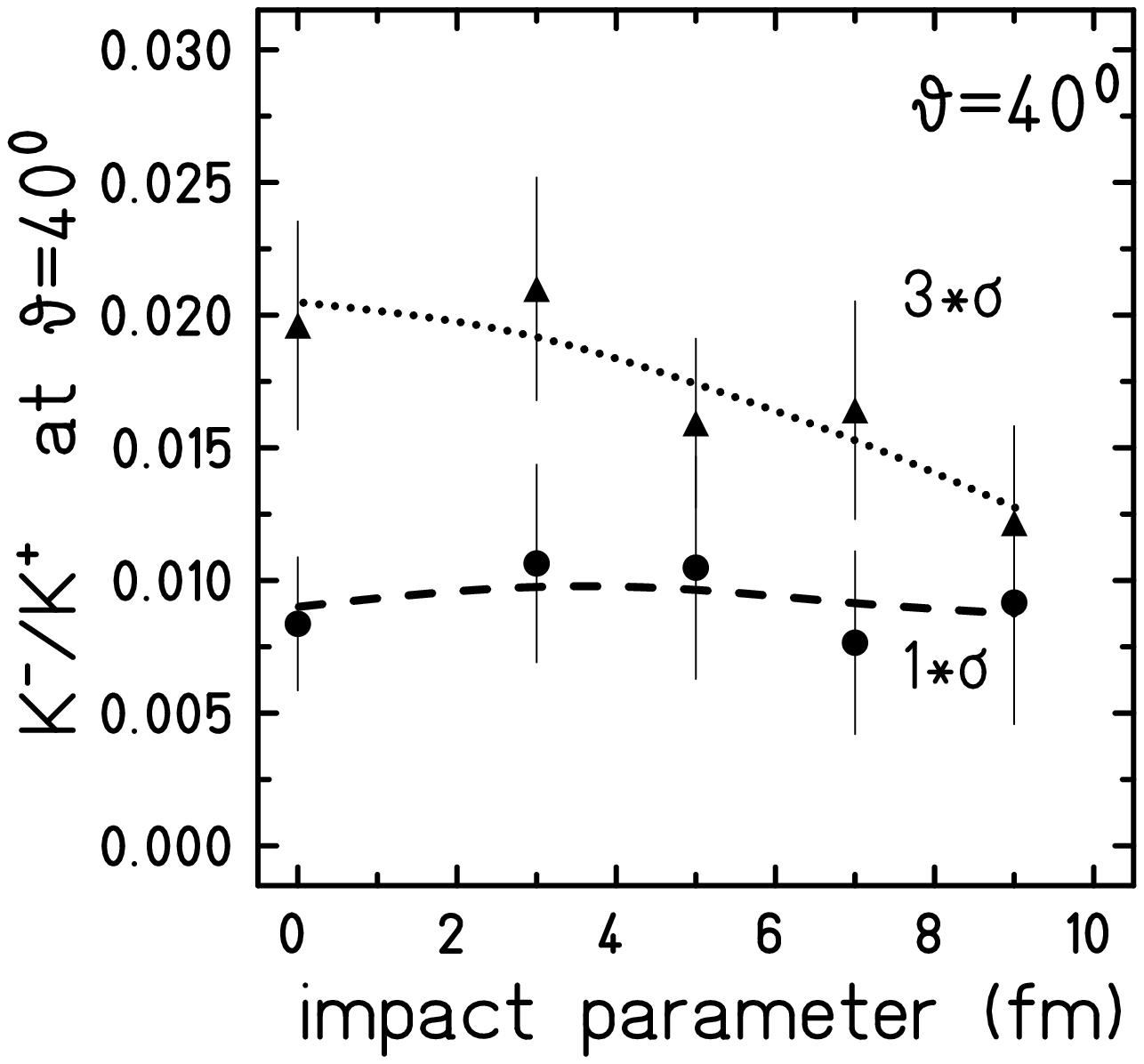,width=0.6\textwidth}
\caption{Impact parameter dependence of the $K^-/K^+$ ratio in $\theta_{lab} =
40 \pm 4 ^\circ$ for Au+Au collisions at 1.5 AGeV. The effective 
cross sections  used in the
calculation are denoted by $\sigma_{eff} = n \sigma_{exp}$ with
$n$ being a multiplication factor.}
\label{impact}
\end{figure}
Next we study how the $K^-/K^+$ ratio depends on the impact parameter.
The results of the IQMD calculations are presented 
in Fig.~\ref{impact}. 
For the standard parametrisation ($n=1$) we observe  
for Au+Au collisions a rather 
constant $K^-/K^+$ ratio for impact parameters smaller than 8 fm in agreement
with the preliminary experimental results \cite{sqm} which confirm the
independence already found in other systems\cite{Barth,Menzel}.
This has been considered as remarkable
because both, the $K^+$ as well as the $K^-$ yield increase with decreasing
impact parameter. 

If the system could be described as a grand canonical ensemble this would be of 
no surprise.
Even in the canonical approach where strangeness is strictly conserved
a constant $K^-/K^+$ ratio is expected
as the terms depending on the system size drop
out~\cite{CLE00}. This observation has triggered the conjecture that
a chemical equilibrium is reached at these beam energies. 
Our calculation shows a microscopic understanding of this impact parameter
independence. We have already seen that the dominant reaction channel for the
$K^-$ is $\Lambda \pi \rightarrow K^- B$. Because the $\Lambda $ is produced
together with the $K^+$, the $K^-$ production is directly coupled to the 
$\Lambda$
density and hence to the number of $K^+$. The
calculations show that the length of the trajectory of the $\Lambda$ in matter
does not change for impact parameters smaller than 8 fm. For larger impact
parameters less $K^-$ are produced, whereas the percentage of reabsorption
remains still almost constant. 
The $K^-/K^+$ ratio depends on the number of pions 
present. The relation between the $K^-/K^+$ ratio and the pion multiplicity
is visible between 1 and 10 AGeV~\cite{ho_s2000}.
Pions are only present in heavy ions reactions and therefore the reaction
mechanism in pp reactions is completely different, where at this energy 
a $K^-$ can only be produced together with a $K^+$.
This explains why the experimental results are that different.  
Already for systems as small as C+C, however,  the pion number is 
sufficient for the $\Lambda \pi$ channel to dominate the $K^-$ production. 
The $K^-$ production stops before the number of pions has reached its 
asymptotic value as can be 
seen in Fig.~1. Therefore, in the $K^-$ production the pion and $\Delta$ 
dynamics is encoded as well. (Please note that we
have not taken into account the small $\Lambda \Delta \rightarrow K^-X$ cross
section \cite{ko1}.)   

Figure 3 exhibits another interesting feature: Increasing ``artificially''
the cross sections (for both, production and absorption) by a factor of 3
one expects naively to get closer to the equilibrium condition, i.e.~a constant
$K^-/K^+$ ratio with impact parameter. However, the opposite is seen: the 
$K^-/K^+$ ratio drops towards peripheral collisions. There are two
reasons for this effect. The first one is related to the increasing amount of
spectator matter in peripheral collisions. In spectator matter $K^-$ can only 
be absorbed but not produced. Increasing both cross sections, the effects of 
absorption becomes more pronounced leading to a decrease of the $K^-/K^+$
ratio for peripheral collisions.
The second reason is connected with the expansion of the system. 
In an expanding system the locally available energy decreases as a function of
time and therefore the endothermic reaction becomes suppressed, as discussed
already in Fig.~2. At the same time, if the cross sections are very different 
for the forward and backward reaction, the reaction with the larger cross section
will continue for a much longer time. Close to threshold (where flux and phase
space are quite different in both directions) both come together and 
therefore the system will run out of equilibrium during the expansion
even if  it has been in thermal equilibrium initially. Hence the final particle 
ratios will be defined by the cross sections.
This phenomenon we see for the case $n$=3 
where a larger $K^-$ absorption which is not 
compensated by a larger production.

Up to now we have studied the $K^-$ production assuming that both, the 
$K^-$ and the $K^+$, have a mass as given by the relativistic mean 
field calculation~\cite{schaf}. These calculations are yet far from being
confirmed by experimental results. It is therefore important to see how the 
predicted mass change of the kaons in the medium influences their multiplicity. 
The $K^-N$ potential is attractive, leading to lower ``in-medium'' masses,
while the $K^+N$ potential is slightly repulsive.
For this reason we study the time evolution 
of the $K^-$ and $K^+$ yields under different assumptions on 
the $KN$ potential: 
We compare the standard calculation 
($K^+ :w, \ K^- :w$,  where w stand for ``with potential'') 
with those in which either the $K^+N$ potential 
($K^+ :w/o, \  K^- :w$) or the $K^-N$ potential ($K^+ :w, \ K^- :w/o$) is 
switched off 
as well as with a calculation in which no $KN$ potential is applied 
($K^+ :w/o, \ K^-:w/o$)and consequently the kaons have their free mass. 

\begin{figure}
\psfig{figure=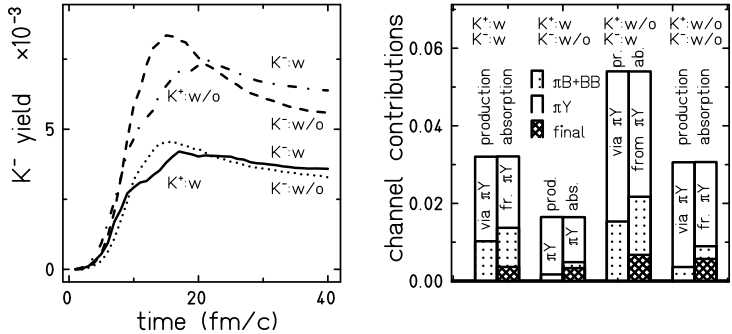,width=0.6\textwidth}
\caption{Left panel: Influence of the $K^{+/-}N$ potentials on the final $K^-$ 
yields
for central Au+Au collisions at 1.5 AGeV. We separately switch off (w/o)  
and on (w) the $KN$ potentials for $K^+$ and $K^-$. For the two upper curves 
the $K^+N$ has been switched off.  Right panel: Production and 
absorption of the $K^-$: right hand column (from bottom to top) the production 
in BB collisions, $\pi B$ collisions and $\pi \Lambda$ collisions, 
left hand column (from bottom to top) survival and absorption.}
\label{potential}
\end{figure}

The result, shown in Fig.~\ref{potential} (left) is evidence that the final $K^-$ 
yield 
depends strongly on the $K^+N$ potential but is almost independent on the 
$K^-N$  potential. This has an easy explanation: the $K^+N$ potential 
determines how many $\Lambda $ are produced in the initial 
$BB\rightarrow \Lambda K^+ B$ reaction. This reaction takes place when 
the baryon density is high. The $K^+N$ potential increases the ``mass'' of the 
$K^+$ and hence their production threshold and lowers therefore the
$\Lambda$ multiplicity. On the contrary, the mass change of the $K^-$ 
has little influence on the result because the observed $K^-$ are created 
very late in the reaction by the mechanism described above and therefore
at a density where the mass change due to the $K^-N$ potential is small.  
Thus heavy ion reactions test the $KN$ potentials at very different densities: 
The $K^+N$ potential is tested around twice nuclear matter density, where the 
$\Lambda$ and $K^+$ are produced, whereas the $K^-$ potential is tested at 
low densities where it is small.  
Furthermore, the final multiplicity of $K^-$ does not depend on the 
$K^-N$ potential  
because the two rates, production and absorption, both depending on the 
potential, balance each other almost completely as both rates depend
on the $K^-N$ potential.
This can be seen from Fig.~\ref{potential} (right) 
where the number of produced and 
absorbed $K^-$ is displayed. As expected, 
the smaller mass of the $K^-$ caused by  the $K^-N$ potential  increases the 
rate (but in both directions 
by about the same amount).  The decrease of the $K^-$ yield at the end of the 
reaction is exclusively caused by the
(exothermic) absorption which still takes place but which is not 
counterbalanced by creation for which the available energy is too small. 

The number of finally observed $K^-$ is directly proportional to
the number of $\Lambda$ produced initially. This number is equivalent to the
number of $K^+$. Therefore we have divided in Fig.~1 the $K^-$ multiplicity by
the $K^+$ multiplicity. We see in Fig.~1 that the ratio $M(K^+)/M(K^-)$ depends 
little on the system size (in distinction to $M(K^+)$).

In conclusion, we have given an interpretation of the experimental observation
that a) in heavy ion reactions the yields for $K^-$ compared to
$K^+$  
is much higher than in pp collisions (compared at the same available energy
with respect to their thresholds) and b) that the $K^+/K^-$ ratio is 
independent of the impact parameter for semicentral and central reactions.
The pair-production channel which is the only channel in pp collisions,
contributes only 
marginally to the finally observed $K^-$ yield in $AA$ collisions.
Almost all $K^-$ are produced by the pionic channel $\Lambda \pi
\rightarrow K^- B$ which is not available in pp. 
Because the $\Lambda$ is produced simultaneously  with 
the $K^+$, the $K^-$ and the $K^+$ production are strongly correlated. 
The naively expected impact-parameter dependence of the $K^-$ yield due to 
absorption is not observed because recreation and absorption occur at almost 
the same rate due to the large difference in the cross section between creation
and absorption.  Despite of the fact that thermal models predict \cite{CLE00}
the $K^-$ multiplicity and the impact parameter independence of the $K^+/K^-$
yield we observe in the simulations that both are determined by dynamical 
quantities like cross sections and by the locking of the $K^-$ to the $K^+$.
The systems are too rapidly expanding for reaching equilibrium in a 
channel which has a relatively small
cross section.  
Furthermore, at the end of the expansion 
the cross section for the exothermic
reaction is large due to the detailed-balance factors.
This is a very generic phenomenon and not
limited to the $K^-$ production and leads the system to run out of equilibrium
during expansion even if it had been in equilibrium before.
The final yield of $K^-$ depend on the $K^+N$ potential (which determines 
how many $ \Lambda $ are produced initially) but does not depend on the 
$K^-N$ potential because the observed 
$K^-$ are produced very late at low densities where this potential is 
small. 

\end{document}